\documentclass[floatfix, preprint, aps, prx]{revtex4-2}

\usepackage{graphicx}
\usepackage{amsfonts, amsmath, amssymb}
\usepackage{bm}
\usepackage{hyperref}
\usepackage[mathlines]{lineno}
\begin{document}

\title{Unraveling the Variations of the Society of England and Wales through Diffusion Maps Analysis on Census 2011}

\author{Gezhi Xiu$^\dag$}
\affiliation{School of Earth and Space Sciences, Peking University}
\affiliation{Center of Complexity Sciences and Department of Mathematics, Imperial College London}
\author{Huanfa Chen}
\affiliation{Centre for Advanced Spatial Analysis (CASA), University College London, London, UK}
\date{\today} 

\begin{abstract}
    We propose a new approach to identifying geographical clustering and hotspots of inequality from decadal census data. We use diffusion mapping to study the 181,408 Output Areas in England and Wales, which allows us to decompose the feature structures of countries in the census data space. Additionally, we develop a new localization metric inspired by statistical physics to uncover the importance of minority groups in London. The results of our study can be applied to other census-like data constructions that include spatial localization and differentiation from low degrees of freedom. This new approach can help us better understand the patterns of social deprivation and segregation across the country and aid in the development of policies to address these issues.
\end{abstract}

\maketitle

\section{Introduction}



Determining the demographic and socioeconomic characteristics that shape a society's overall picture is crucial. The distribution of social classes and groups within a society can greatly impact the region's stability, social welfare, and economic potential~\cite{holzer2003public,yang2022oversampling,arva2016spatial}. For example, \cite{barter2019manifold} has shown that, for some cities, university students and social deprivation are the most significant factors in explaining other social variables in census statistics. Other studies have revealed the impact of social, economic, and ethnic attributes on regional disparities, such as energy burdens in households~\cite{scheier2022measurement}, heterogeneity in epidemic vulnerability~\cite{elliot2000spatial}, and environmental inequality~\cite{brazil2022environmental}.

Census data or similar datasets are often used to identify these demographic characteristics by grouping a large number of social variables collected from hundreds of thousands of geographical units (thus, country-level) into a few independent spatial distributions~\cite{ratledge2022using}. However, due to the scale and complexity of these datasets, there can be challenges in practice. These may include limitations in processing only a selection of social variables on a large spatial scale~\cite{liu2019principal}, or utilizing a full collection of social variables for only a small region with expert knowledge~\cite{liu2019principal}.

We argue that the objectivity can be compromised when deriving these social variables or spatial regions from census data. For instance, defining geographical clustering of social groups for statistical analysis requires aggregating regions into specific areas. However, the modifiable areal unit problem (MAUP)~\cite{gehlke1934certain} challenges the possibility of such spatial aggregations being consistent across different social issues. Additionally, the heterogeneity of features across cities leads to a gap between local studies and the general significance of these features for all cities. Furthermore, researchers may have different perspectives on nominal attributes like race or religion, leading to a lack of consensus on the significance of these features. This can make it difficult to synthesize findings from different studies to identify critical socioeconomic characteristics.

To overcome these challenges, the Diffusion Maps (DM) manifold learning method can be applied to analyze census data. DM effectively captures the interplay of social indicators by assigning the social identities of locations as branches in the data manifold. Previous studies, such as~\cite{barter2019manifold}, have used DM to analyze similar cities and identify common social indicators, decomposing 1450-dimensional census data into the two most significant social indicators of Bristol and Edinburgh (university students and social deprivation). In fact, it forms a successful attempt to overcome the MAUP problem by defining a topic-specific continuous metric that is localized into small patches of important areas of census.

While this approach has been successful in identifying important social indicators in Bristol and Edinburgh, it may not accurately represent the entire population of England and Wales. To gain a comprehensive understanding of demographics, it is important to identify globally-consistent contributors while also taking into account the unique characteristics of small communities within a society. To achieve this, our study applies DM to the census data of England and Wales to identify geographical clustering and hotspots of inequality, providing a more nuanced understanding of demographics.

Our study utilizes Diffusion Maps to analyze the census data of England and Wales and identify geographical clustering and hotspots of inequality. The goal is to decompose high-dimensional social variables into branched, interdependent social factors, revealing patterns in space that would otherwise go unrecognized. A new method, the correlation table, is proposed to explain the derived social dynamics and provide a standard for structuring and analyzing any spatial collection of features while minimizing preassumed spatial correlations in large study areas, such as a densely organized country. Additionally, we introduce a localization metric to reveal the critical features of specific cities. Our method provides a comprehensive view of the descendingly important features of England and Wales and tracks where these features are locally highlighted. By combining the Diffusion Maps method with the correlation table and localization metric, our study offers a powerful tool for understanding demographics and uncovering patterns in social and economic data.

\section{Method and Data}

\subsection{The census data}

The 2011 UK Census collected by the Office for National Statistics of the United Kingdom (https://www.ons.gov.uk/census) provides a comprehensive picture of the population and households in England and Wales, with over 2,000 social variables (or features) on 181,408 Output Areas (OAs), where OAs are the smallest geographical units used in the census, are designed to be compact, homogeneous, and contain between 125 and 650 households. However, the vast amount of data can make it difficult to extract meaningful insights. Our study aims to tackle this challenge by utilizing Diffusion Maps to decompose the high-dimensional social variables into branched social factors, revealing hidden patterns in space. 


\subsection{Diffusion Maps}


Diffusion Maps (DM) is a nonlinear dimensionality reduction technique that leverages a random walk process on a sparse network of data points to uncover the structural differentiation within data. In urban sciences, it is sensible from the idea that locations are clusters of similar individuals. Hence, the similarity of locations is equivalent to their distances to each other in the data space, which can further be used to define the network topology. This method allows for a local perspective to be integrated into a broader understanding of urban dynamics, making it an ideal tool for our study.

The constructions of the DM are performed as follows. Suppose for each of the $N$ Output Areas, $x$ is a $M$-dimensional vector whose entries are the social variables. Here, $M$ is the dimensionality of social variables in the census dataset, and $N$ is the total number of data points. To leverage the distribution heterogeneity of different social variables, we measure the similarity $s(x, y)$ of the OA pair $x$, $y$ through their Spearman Rank Correlation 
\begin{equation*}
\rho_{x,y}=\frac{R_x R_y - N^2/4}{||R_x-n/2||\cdot ||R_y-n/2||}
\end{equation*}
for each pair $(x, y)$ in $\{1,\dots, N\}$, where $R_x$ is a vector that each of its entry is the rank of $x$ for a social variable. We denote $\Sigma$ as the rank correlation matrix, where each of its elements $\Sigma_{x,y} = \rho_{x,y}$ is the correlation of the corresponding data points $x$ and $y$. The elements of $\Sigma$ are thus all valued between $-1$ and $1$. Nearby points in the data space have $\rho_{x,y}$ close to 1 following a framework in \cite{ryabov2022estimation}. To emphasize the structure of the most important links connecting most similar data points, we define an alternative matrix $\widetilde{W}$ keeps only $k$ largest elements in each row of $\Sigma$ and lets the rest of elements be zero. Here, we choose $k=10$ that barely keeps the network connected, that from each data point there exists at least one route to every other data point in the network. Next, we define a $N\times N$ normalization matrix $D$ whose diagonal elements are the row sums of $\Sigma$. Then, we compute the eigenvalues and right eigenvectors of the following normalized Laplacian matrix $A = I - D^{-1}\widetilde{W}$. $A$ can be regarded as a Markovian transition matrix for a random walk process over data points. The random walk process converges to a continuous time diffusion process as $N\to \infty$ and a small $k$ over the observable data manifold. The low-order eigenvectors of $A$ are then an approximate parameterization of the underlying manifold that hints at the actual urban dynamics.

As presented in~\cite{barter2019manifold}, the social features can be represented by the linear combination of the leading eigenvectors. The complete set of eigenvectors $\eta_j$ correspond to an increasing sequence of $A$’s eigenvalues $\lambda_0 \le \lambda_1 \le \dots \le \lambda_{N-1}$, and each of $\eta_j$ corresponds to a relatively independent dynamical variable, whose nonlinear combinations are explicit in the census as social variables. We then color code the output areas according to their corresponding elements in each eigenvector $\eta_j$, and generate spatial plots to visualize the spatial configurations of the dynamical variables. 

In order to make sense of the dynamical variables identified through the eigenvectors, we perform a backward calculation to investigate the correlation between the eigenvectors $\eta_j$ and the census social variables. By identifying the social variables that are most positively and negatively correlated with a given eigenvector $\eta_j$, we can gain valuable insights into the significance of the corresponding eigenfeatures. This information, in combination with the visual representation of the eigenvectors through spatial plots, enables a comprehensive analysis of the underlying dynamics. 

\subsection{Virtual similarity networks versus social hierarchy}

From the above analysis of DM, we are actually interpreting the census dataset as a weighted sparse network formed by 181,408 Output Areas (OAs) in a 1,450 dimensional space, and the `similarity' of OAs as weights of the links. The sparsity of this census data network is required to extract the backbones of the feature's synthetic structure rather than exposing to some highly heterogeneously distributed social variables. A natural extension of the problem is then how sparse the network needs to be to recover the underlying structures and properties of the census data, represented by the social hierarchy and criticality~\cite{tadic2017mechanisms}. Consequently, the discussion on whether there are significant cross-scale features in the census data helps to justify the proposed network based on local metrics.

We here consider the problem of sparsity by simply assuming the census data network should be at a most informative criticality state because the society is widely accepted to be a critical system~\cite{midgley1994ecology}. 

As the census data network is formed by finding the $k$-most similar OAs for each of the OAs, the sparsity of the network can then be determined through the value of $k$: a larger (smaller) $k$ represents stronger (weaker) network connectivity, and denote $G_k$ as the network of connectivity $k$. 
To define the census data network's criticality, we specifically consider the degree distribution of the census data network for different network sparsity~\cite{larremore2011predicting}. Table~\ref{tab:powerlaw_diff_map} shows that as the similarity network $G_k$ is defined as more sparse (i.e., smaller $k$), the likelihood of $G_k$'s degree distribution being more similar to a powerlaw distribution increases. This suggests that as the adjacency threshold and network connectivity decrease, the network's powerlaw characteristics become more prominent. The analysis in this paper thus chooses $k=10$ to maximize the likelihood for the data network to be powerlaw-like. 

\begin{table}[t]
    \centering
    \begin{tabular}{|c|c|c|c|c|}
    \hline
        $n$ & $\alpha$ & $x_{\min}$ & $R$ & $p$ \\ \hline
         10 & 2.890    & 16.       & 0.933 & 1.846$\times 10^{-9}$\\ \hline
         20 & 2.877    & 30.       & 0.0385 & 9.367$\times 10^{-1}$\\ \hline
         30 & 2.878    & 44.       & -2.726 & 1.381$\times 10^{-1}$\\ \hline
         40 & 2.875    & 58.       & -9.835 & 4.938$\times 10^{-3}$\\ \hline
         50 & 2.867    & 73.       & -27.09 & 3.493$\times 10^{0}$\\ \hline
    \end{tabular}
    \caption{The maximum-likelihood fitting methods combined with goodness-of-fit tests (based on the Kolmogorov-Smirnov statistic and likelihood ratios) of $G_k$'s degree distribution. Columns $\alpha$ and $x_{\min}$ represent the estimation of the powerlaw degree distribution of the census network of OAs with the form \(p(k)=k^{-\alpha}/\sum_{n=1}^{\infty} ( n+x_{\min} )^{-\alpha}\), where $x_{\min}$ is the minimum degree, and $\alpha$ is the powerlaw exponent. $R$ is the likelihood ratio test comparing the fit of the power laws curve and the log-normal curve. A more positive $R$ indicates a better fit of the powerlaw curve to the degree distribution over the log-normal curve. Finally the $p$ column is the $p$-value of the confidence of the powerlaw distribution.}
    \label{tab:powerlaw_diff_map}
\end{table}

\subsection{Localized participation ratio}

The DM eigenvectors are globally consistent features that play a significant role in the distribution of a variety of social variables found in census data. There are likely many factors that contribute to the society of England and Wales, and while the most dominant factors may be important, less dominant ones can also have significance in specific regions. This is illustrated in the example of Bristol and Edinburgh where university students have higher socioeconomic importance than social deprivation.

Here, we emphasize the importance of understanding how certain factors that are important on a global scale (as represented by the leading DM eigenvectors) are localized in specific cities. To accomplish this, we introduce a new metric called the local inverse participation ratio ($LIPR$) which allows us to trace the localization of an eigenfeature (i.e. a specific factor represented by an eigenvector) into a certain city. They argue that this is important because it can help to identify cities that are of special importance for a particular factor. Additionally, we note that there are few metrics in existing literature that measure the local properties of global features and therefore it is essential to introduce a new localization index like $LIPR$.

The $LIPR$ is an extension from the metric `inverse participation ratio' (IPR) from statistical physics~\cite{fyodorov1992analytical}, defined as~\begin{equation}
    IPR_i = \sum_{k=1}^M \frac{(\eta_i^k)^4}{\sum_{j=1}^{M} (\eta_i^j)^2},
\end{equation} where $M$ is the number of $\eta_i$'s entries, and $\eta_i = (\eta_i^1, \dots, \eta_i^M)^T$. Here, if a feature appears in one single area, i.e., $\eta_i = (0,\dots,0, 1, 0, \dots, 0)^T$, the corresponding $IPR_i = 1$; for another limiting case, if a feature is uniformly distributed in all the areas, $\eta_i = (1/\sqrt{M},\dots, 1/\sqrt{M})^T$, the corresponding $IPR_i = 1/M$, which diminishes as $N$ grows. So a highly localized pattern corresponds with a large value of the $IPR$.

To capture whether an indicator clusters in an area, we extend the $IPR$ to local inverse participation ratio $LIPR$ of area $A$,~\begin{equation}
    LIPR_i^A = \left( \frac{\sum_{j\in A} {\eta_i^j}^4}{||\eta_i||^2}\right) / \left( \frac{\sum_{j} {\eta_i^j}^4}{||\eta_i||^2}\right) = \frac{\sum_{j\in A} {\eta_i^j}^4}{\sum_{j} {\eta_i^j}^4}.
\end{equation} It is intended to be large when the distribution of eigenfeature $i$ is highlighted in the city $A$. A region with a high $LIPR$ indicates the spatial clustering of small communities, which supports similar social groups across the country, and is mainly localized in some single cities. 

The $LIPR$ metric can be used to understand how localized an eigenfeature is in a certain city. We give two examples to illustrate how the metric works in two limiting cases. In the first case, an eigenfeature highlights only one area in London and assigns it a value of 0.1, while assigning 0 to all other areas. The corresponding $LIPR$ in this case would be a relatively high value of 0.001. In the second case, if an eigenfeature does not highlight any specific areas in London and assigns all 10,000 areas a value of 0.0001, the corresponding $LIPR$ would be a near-zero value of $10^-8$. They explain that in general, a highly localized eigenfeature would have a larger $LIPR$ value and that the metric can be used to pinpoint meaningful communities in more than one city.

\section{General dominant features}

We begin at the smallest positive, thus the most important Laplacian eigenvectors of the England and Wales diffusion mapping. A spatial plot can associate each of the eigenvectors, which is color-coded from the most negative to the most positive entries, representing the exposures of each OA to the corresponding demographic context.

\subsection{Urbanization properties}

\begin{figure*}[ht]
    \centering
    \includegraphics[width=0.9\linewidth]{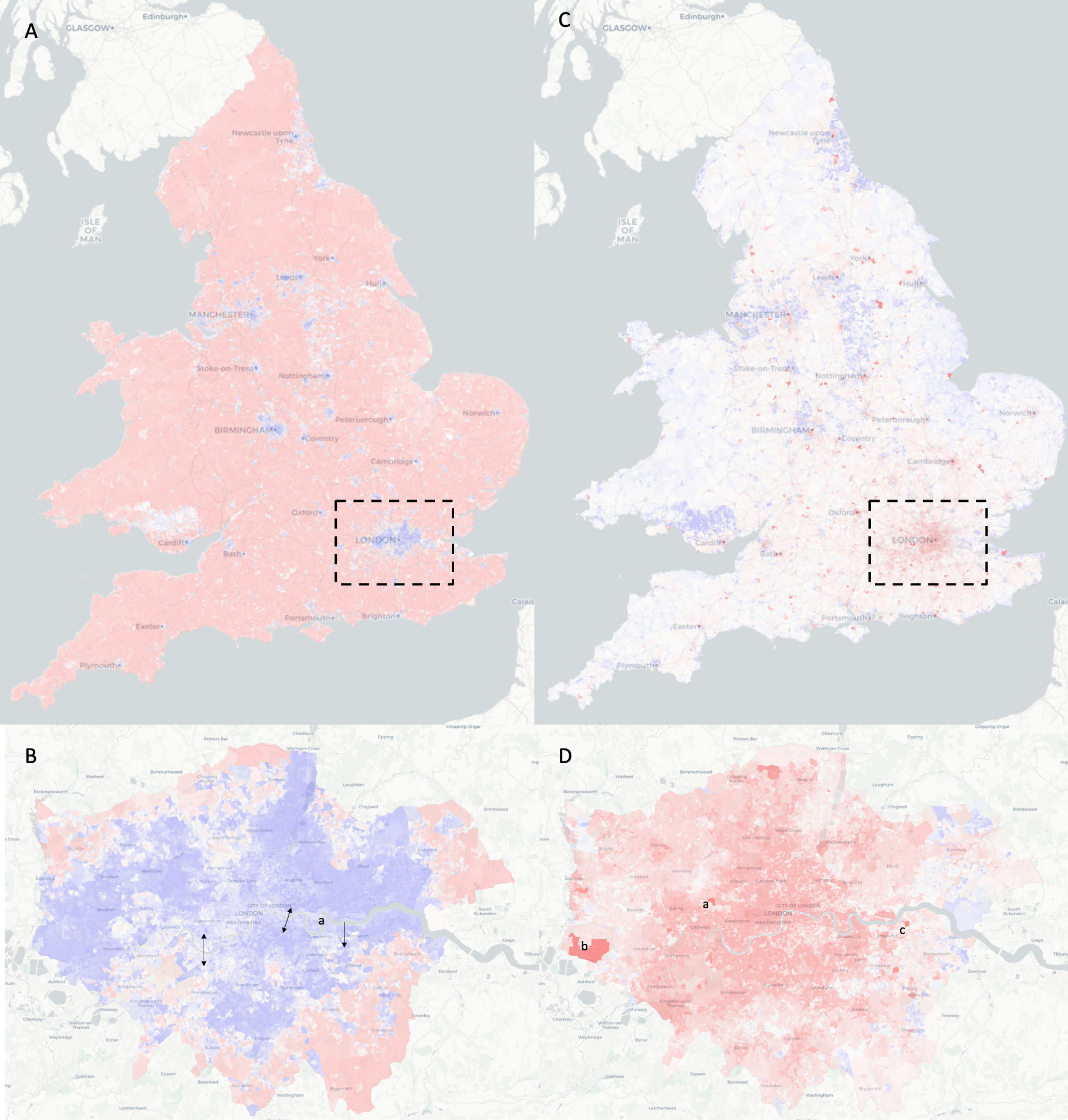}
    \caption{The eigenvector map of $\eta_1$ and $\eta_2$, the representative eigenvectors that shows global properties. The colors are assigned by the entries of an eigenvector, from the most positive (red) to the most negative (blue). Here, $\eta_1$ highlights the main cities; $\eta_2$ pinpoints the most skilled workers, which is mostly concentrated in the main airports. The label \textbf{a} in \textbf{B} is the Blackwell tunnel from where the symmetric pattern is broken between either sides of Thames; In \textbf{D}, \textbf{a}, \textbf{b}, and \textbf{c} are Hammersmith Hospital, Heathrow Airport, and HM Prison Isis, respectively.}
    \label{fig:EF12}
\end{figure*}
The first eigenvector, $\eta_1$ can be used to identify patterns of urbanization in England and Wales~(Fig.~\ref{fig:EF12}\textbf{A} and \textbf{B}). $\eta_1$ is localized in the main cities of the country, and that it highlights not only the largest cities such as London, Liverpool, and Manchester, but also smaller central places surrounded by forest and mountains in the form of a continuous patch of OAs represented by Porthmadog, Tregaron, and Newport. By analyzing only London entries of $\eta_1$, we find working-class residential areas expanding along the River Thames, with a relatively north-south symmetrical pattern from west to east until the Blackwall tunnel neighborhood, where tunnels replace the walkable bridges as the connection between the riversides. We conclude this as urbanized residential areas are the walkable neighborhood, which is the most explanatory feature of the 2011 census. We recall the diffusion mapping results inputting the city-level census data of in~\cite{barter2019manifold} that highlight universities and poverty as the dominant features of Bristol and Edinburgh. The eigenvector $\eta_1$ exhibits a more globalized spatial distribution of urbanization. 

A natural question to follow is what element from the census perspective determines the shape of a city identified by $\eta_1$. To this end, we compute the correlation of $\eta_1$ with all the census social variables. We found that the most correlated census variables of $\eta_1$ and the corresponding correlations are: \textit{living in a couple: Married or in a registered same-sex civil partnership} (0.82), \textit{two cars or vans in household} (0.80), \textit{Married} (0.78), \textit{Occupancy rating (rooms) of +2 or more}, i.e. at least 2 rooms more than the basic standard (0.78); Meanwhile, $\eta_1$ is also highly negatively correlated with particular races and religions (\textit{Black African/Caribbean/Black British: African} -0.58, \textit{Muslim} -0.57). These social variables capture the typical community in a city in England and Wales. We note that urbanization is the most important dimension in census, and urbanization is largely explained by the percentage of marriage and civil partnerships, vehicles ownership, and occupancy status of the households in a neighborhood.

Eigenvector $\eta_2$ highlights similar areas $\eta_1$ but exhibits a milder aggregation with many clustered areas in medium level regional centres~(Fig.~\ref{fig:EF12}\textbf{C} and \textbf{D}). Generally, $\eta_2$ picks all the important airports in England and Wales with the highest entries, in addition to a general mapping of the working class in most cities and lower-level central places. We conclude that $\eta_2$ is mostly associated with the skilled occupations, which can also be validated statistically by its most correlated census variables of degrees and diplomas: \textit{Degree (for example BA, BSc), Higher degree (for example MA, PhD, PGCE)}, 0.85, \textit{two+ A levels/VCEs, 4+ AS levels, Higher School Certificate, Progression/Advanced Diploma, Welsh Baccalaureate Advanced Diploma} (0.81), \textit{Highest level of qualification: Level 4 qualifications and above} (0.81). $\eta_2$'s high correlation with education and its appearance at the second most dominant eigenvector indicate that education is one of the most clustering feature of England and Wales, that widely explains other socioeconomic properties underlying census data.

We then wonder what areas are `most educated'. Zooming in on London, $\eta_2$ separates the city from Northwestern to Southeast, similar to what is usually believed as the separation of Old and New London. The most highlighted areas of $\eta_2$ in London are the Hammersmith Hospital. However, $\eta_2$ surprisingly finds HM Isis Prison. We referred to the prison website and Wikipedia and learned that this prison provides education and vocational training in partnership with Kensington and Chelsea College. 

Beyond educations, $\eta_2$ is highly negatively correlated with \textit{Routine occupations} (-0.75), \textit{No British identity} (-0.60), and \textit{Bad health} (-0.59). These features indicate that education is one of the most important determinations of household gathering features as the education-related eigenvector appears to be as $\eta_2$. Here we compare the spatial distribution of $\eta_2$ and $\eta_6$ because visually $\eta_6$ finds almost every university in England and Wales. We conclude that $\eta_2$ is more about where the university graduates settle and work, while the positive entries of $\eta_6$ find most of the university campus. The population composition of $\eta_6$'s most correlated with the racial census variables are \textit{White: English/Welsh/Scottish/Northern Irish/British} (0.57), \textit{No religion} (0.55), and \textit{Born in UK} (0.53). These features can be linked to the typical features of the university neighborhood of England and Wales.

\subsection{University neighborhoods}

\begin{figure}[ht]
    \centering
    \includegraphics[width=0.99\linewidth]{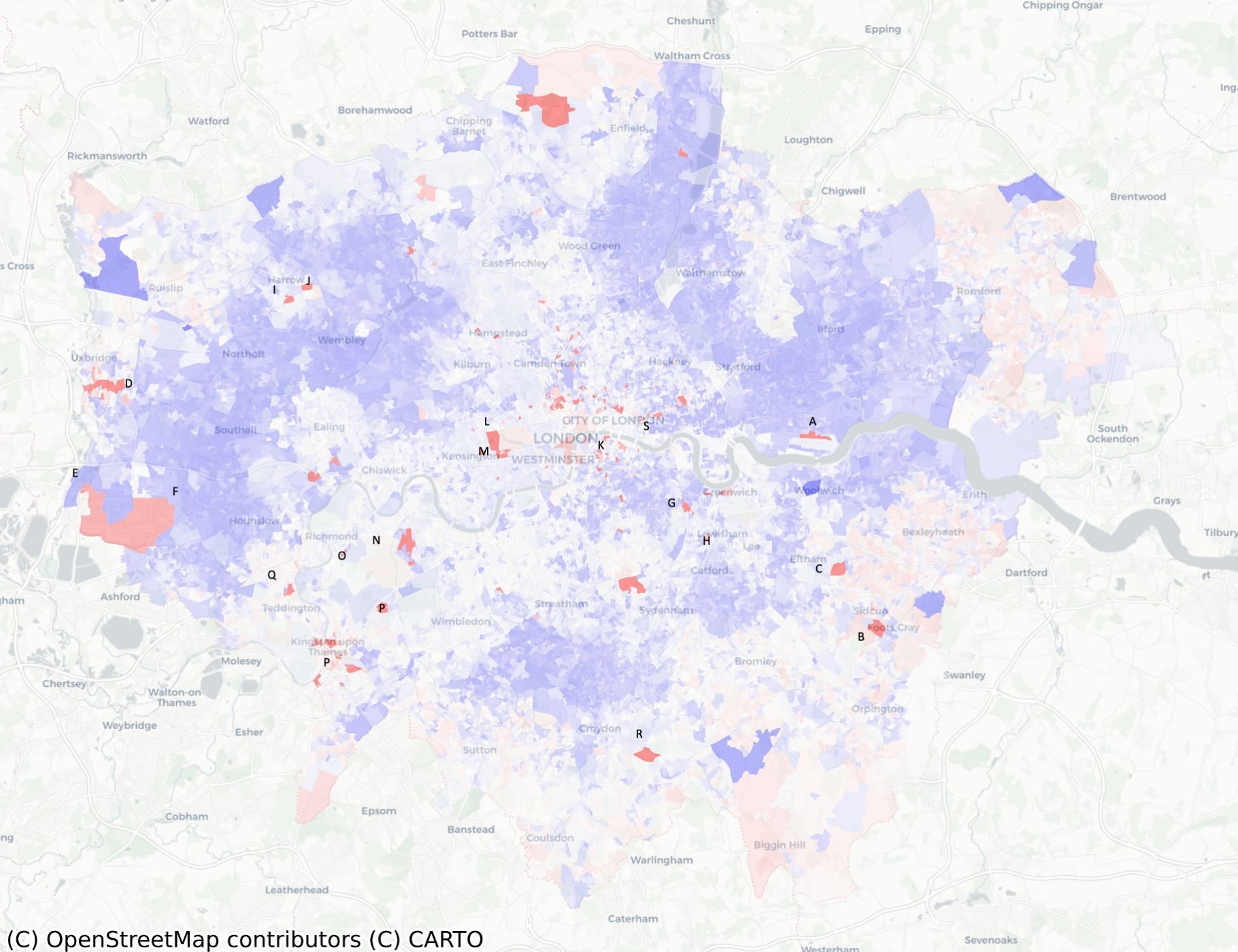}
    \caption{The eigenvector $\eta_6$ zoomed in London, which finds A. University of east London, B. Queen Mary’s Hospital, C. University of Greenwich, D. Brunel University, E. Smart College UK, F. The London College, G. Goldsmiths, University of London, H. Lewisham College, I. Harrow School, J.Northwick Park Hospital, K. King's College London Guy's Campus, L. Imperial College, M. Chelsea and Westminster Hospital, N. University of Roehampton London, O. Richmond University, P. Kingston University, Q. St Mary's University Twickenham London, R. Croydon College, S. Northumbria University - London etc.}
    \label{fig:london6}
\end{figure}

The spatial pattern of variable $\eta_6$ is associated with universities, which is not expected to be related to ethnicities. However, statistical analysis reveals differences in correlation with various ethnicities. The correlation coefficient between $\eta_6$ and the ethnic group of \textit{White: English/Welsh/Scottish/Northern Irish/British} is high at 0.570, while it has a negative correlation with \textit{British only identity}, \textit{self-employed individuals}, and the \textit{African language group of Somali}. These correlations are likely due to historical factors, as universities were established at a time when fewer immigrants came to the UK for education, and university communities tend to be selective or stable, with many graduates having a strong emphasis on education and research.

At a finer level of correlation, $\eta_6$'s correlation with individuals who identify as having \textit{No Religion} is 0.546. This can be explained by the high proportion of non-religious researchers in scientific or social studies, as well as the high proportion of non-religious international students in university-related areas. Other social variables that have correlations with $\eta_6$ that are greater than $0.5$ include \textit{Born in the UK} (0.530), \textit{Europe: Total} (0.508), and \textit{No British identity} (0.502). Census data was collected at the household level to identify households with pure British or foreign backgrounds in the highly correlated social variables. This household composition is representative of the typical characteristics of university staff and students, including middle-aged families established prior to recent globalization and young students in shared tenancy arrangements.

\subsection{Social security: prisons and military installations}

Eigenvector $\eta_3$ was found to have a high correlation with prison installations, as evidenced by its strong association with the social variable \textit{Other establishment: Prison Service} and \textit{Other establishment: Detention Centers and other detention} (correlation coefficient valued 0.855). This correlation suggests that areas with similar population compositions to prisons are characterized by a unique pattern that may reflect societal instability.

To further validate this association, we examined the correlations of other social variables with $\eta_3$. Our analysis revealed that several factors, including race, education, and health, contribute to an area's stability. Specifically, we found that $\eta_3$ was positively correlated with White: English/Welsh/Scottish/Northern Irish/British (correlation coefficient valued 0.128), \textit{No qualifications} (0.125), \textit{Routine occupations} (0.124), \textit{Fair health} (0.107), and \textit{Last worked before 2001} (0.103).

Of these social variables, health was found to have a particularly interesting relationship with $\eta_3$. Our analysis showed that \textit{medium health conditions}, rather than \textit{very good}, \textit{good}, \textit{bad} or \textit{very bad health}, were mainly related to $\eta_3$. This result is intuitive as individuals in perfect health are likely to have adequate income and those in poor health are less likely to commit a crime. Taken together, these findings provide further support for the hypothesis that $\eta_3$ is a marker of societal instability, and suggest that the distribution of population characteristics related to race, education, and health may play a role in shaping the spatial pattern of crime and prison. These implications are useful for policymakers and researchers seeking to understand and address the root causes of instability in society. 

\section{Feature localization into cities}

\begin{table*}[ht]
\tiny
\begin{tabular}{|c|c|c|}
\hline
$\eta_i$ & $LIPR$ & Interpretation \\
\hline
3  & 0.010373 & Prison service \\ \hline
8  & 0.011114 & Educational establishment  \\ \hline
4  & 0.022512 & Defense establishment \\ \hline
9  & 0.026599 & Retirement \\ \hline
10 & 0.030878 & Defense \\ \hline
7  & 0.058357 & Full time employee \\ \hline
5  & 0.064503 & One person household/Household spaces with no usual residents -\textgreater Tourist? \\ \hline
15 & 0.066020 & Multi-person household: All full-time students averaged household spaces\\ \hline
6  & 0.077067 & University \\ \hline
0  & 0.138103 & -  \\ \hline
12 & 0.143874 & 1 car or van in household/Lower supervisory and technical occupations \\ \hline
14 & 0.179744 & Owned: Owned with a mortgage or loan/Economically active: Employee: Full-time \\ \hline
19 & 0.261445 & One family only: Married or same-sex civil partnership couple: All children non-dependent/ \\ & &  Other households: Three or more adults and no children \\ \hline
17 & 0.264369 & Intermediate occupations/Multiple types of central heating \\ \hline
1  & 0.274687 & Marriage/many vehicles/redundant rooms \\ \hline
11 & 0.295386 & Skilled trades occupations/Lower supervisory and technical occupations/ \\ & & Caring, leisure and other service occupations \\ \hline
2  & 0.303946 & Higher degree -\textgreater Finance and technology \\ \hline
16 & 0.567684 & Tamil/Opposite: Yiddish/Israeli \\ \hline
18 & 0.716807 & Gas central heating/Solid fuel \\ \hline
13 & 0.743838 & Gas central heating, three or more adults and no children, highest level of education \\ \hline
\end{tabular}
\caption{The ranked $LIPR$s of eigenvectors restricted in London. The larger $LIPR$ of an $\eta$ indicates that the feature is more localized in London. Generally, the features with small $LIPR$s are infrastructures, while the features with larger $LIPR$s are the superlinear urban indicators.}
\label{tab:lipr}
\end{table*}

Diffusion maps are a useful method for evaluating and aggregating social variables across different regions. While a variable may hold global significance, it can still hold crucial importance in comprehending the behavior of specific cities if it displays unique characteristics within those areas (such as universities in \cite{barter2019manifold}). These maps can help identify key characteristics of specific areas and uncover minority groups that are concentrated in specific communities within a city. These communities may act as country-wise hubs for that group, e.g., the London Chinatown for Chinese group in London and surrounding areas. To highlight these unique features, we introduce the concept of Localized Inverse Participation Ratio ($LIPR$), which is described in detail in Method section.

\begin{figure*}[ht]
    \centering
    \includegraphics[width=\linewidth]{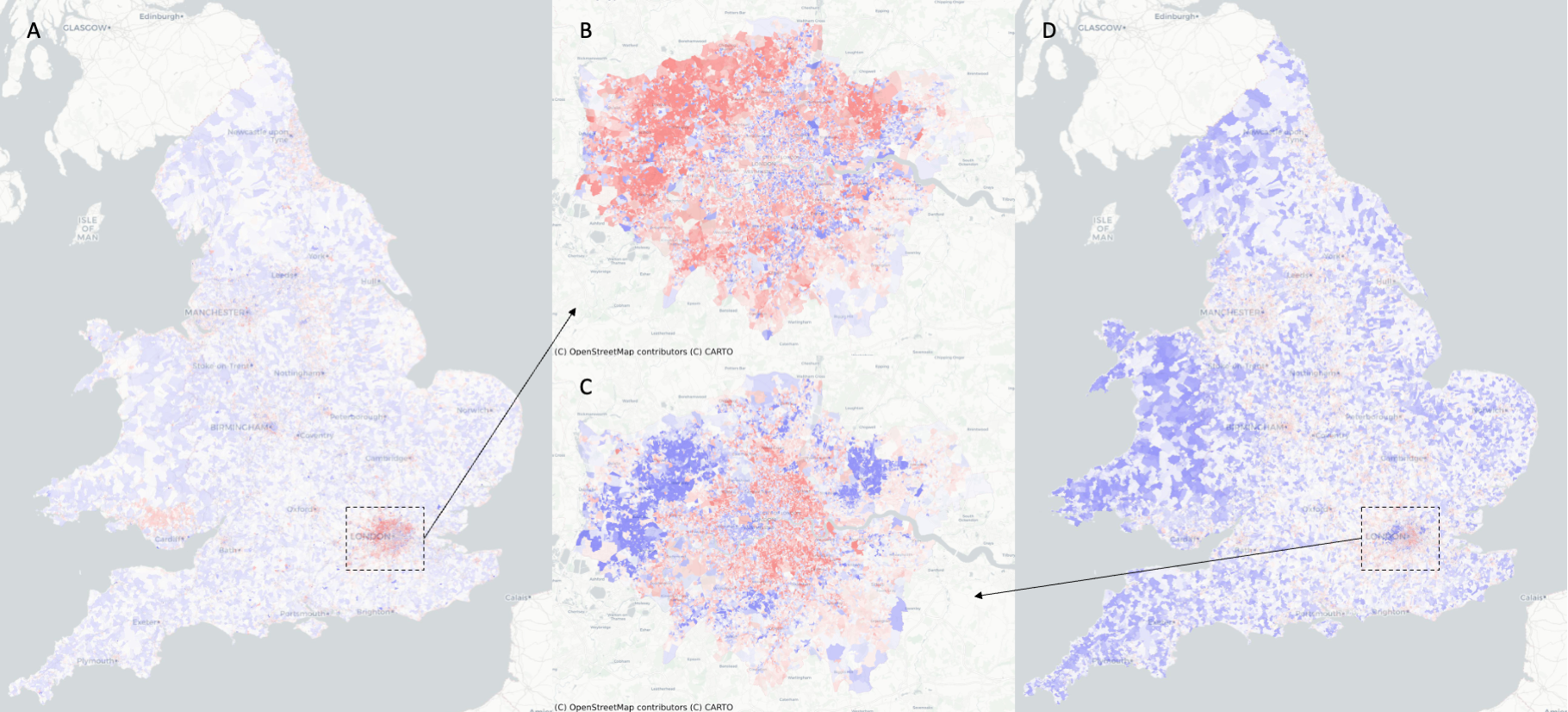}
    \caption{The most localized features in London, $\eta_{13}$ and $\eta_{18}$ for typical lifestyles. Here \textbf{A} and \textbf{B} are the spatial map of $\eta_{13}$ for England and Wales, and London, respectively; \textbf{D} and \textbf{C} are the spatial map of $\eta_{18}$ for England and Wales, and London, respectively.}
    \label{fig:local_feat}
\end{figure*}

We investigate London to show how $LIPR$ is used. First we determine the study set of the first 20 eigenvectors, to pinpoint some of the properties that are important aggregation of social variables valid for the whole England and Wales. Then for each of the eigenvectors, we query the entries that correspond to the OAs in London and further compute the $LIPR$ for the eigenvector-city pair~\ref{tab:lipr}. A benchmark for $LIPR^\text{London}$ values is the uniform distribution, where a feature takes the same value of $1/\sqrt{N}$ in all the OAs in England and Wales, where $N=181,408$. In the Greater London region, there are $N_\text{London}=24,927$ OAs, and the corresponding ``neutral'' $LIPR$ value is $LIPR^*=24927/181408=0.137408$. For an eigenvector $\eta_i$, if its $LIPR_i^\text{London}$ is greater than $LIPR^*$, it can be referred to as a \textit{localized} feature in London; otherwise, if $LIPR_i^\text{London}$ is smallest than $LIPR^*$, $\eta_i$ is not a localized feature in London (either not localized at all, or localized in other cities). A localized feature in London refers to a unique and distinguishable community that is highly concentrated within the city of London, setting it apart from its surrounding neighborhoods. Specifically, if an eigenvector has a high inverse participation ratio (IPR) but a low $LIPR$, it means the corresponding feature is globally significant but not localized in the city. On the other hand, if a feature (such as prisons) has a high IPR and a low LIPR of a city, the feature usually correspond to those rarely seen but essential elements for every city thus infrastructures.

The $LIPR$-ranking approach allows for a systematic investigation of the small social groups in the country who gather in London and have a significant impact. The top localized features in London are $LIPR_{13} = 0.74$, $LIPR_{18}=0.72$, $LIPR_{16}=0.56$, $LIPR_2=0.30$, $LIPR_{11}=0.29$, $LIPR_1=0.27$, $LIPR_{17}=0.26$, $LIPR_{19}=0.26$, $LIPR_{14}=0.18$, and $LIPR_{12}=0.14$ (in descending order). The rest of the eigenvectors may not be localized in London, but could be localized in other cities. 

The correlation analysis of localized features in London reveals that $\eta_{13}$ and $\eta_{18}$ (Fig.~\ref{fig:local_feat}) are highly associated with central gas heating, highest level of education, and households consisting of three or more adults with no children. The correlation coefficients of these features with $\eta_{13}$ and $\eta_{18}$ are around 0.25, which highlights the demographic composition of the typical Londoner. Central heating is more prevalent in newer and more expensive homes, and these homes are more likely to be occupied by higher-educated and childless individuals. The concentration of such households in the affluent suburbs of London is consistent with the trend of urban gentrification and high demand for modern and comfortable living environments in urban areas. Our findings suggest that this demographic is characterized by well-educated individuals living in new build properties with central gas heating. The spatial distribution of $\eta_{13}$ highlights the affluent suburbs of London, which suggests that this area is perceived as desirable by wealthy families in business. This finding is supported by previous research studies~\cite{hanmer2017actors}, which have demonstrated a positive relationship between education level, household composition, and central heating system with wealth and urban development.

\begin{figure}[ht]
    \centering
    \includegraphics[width=0.99\linewidth]{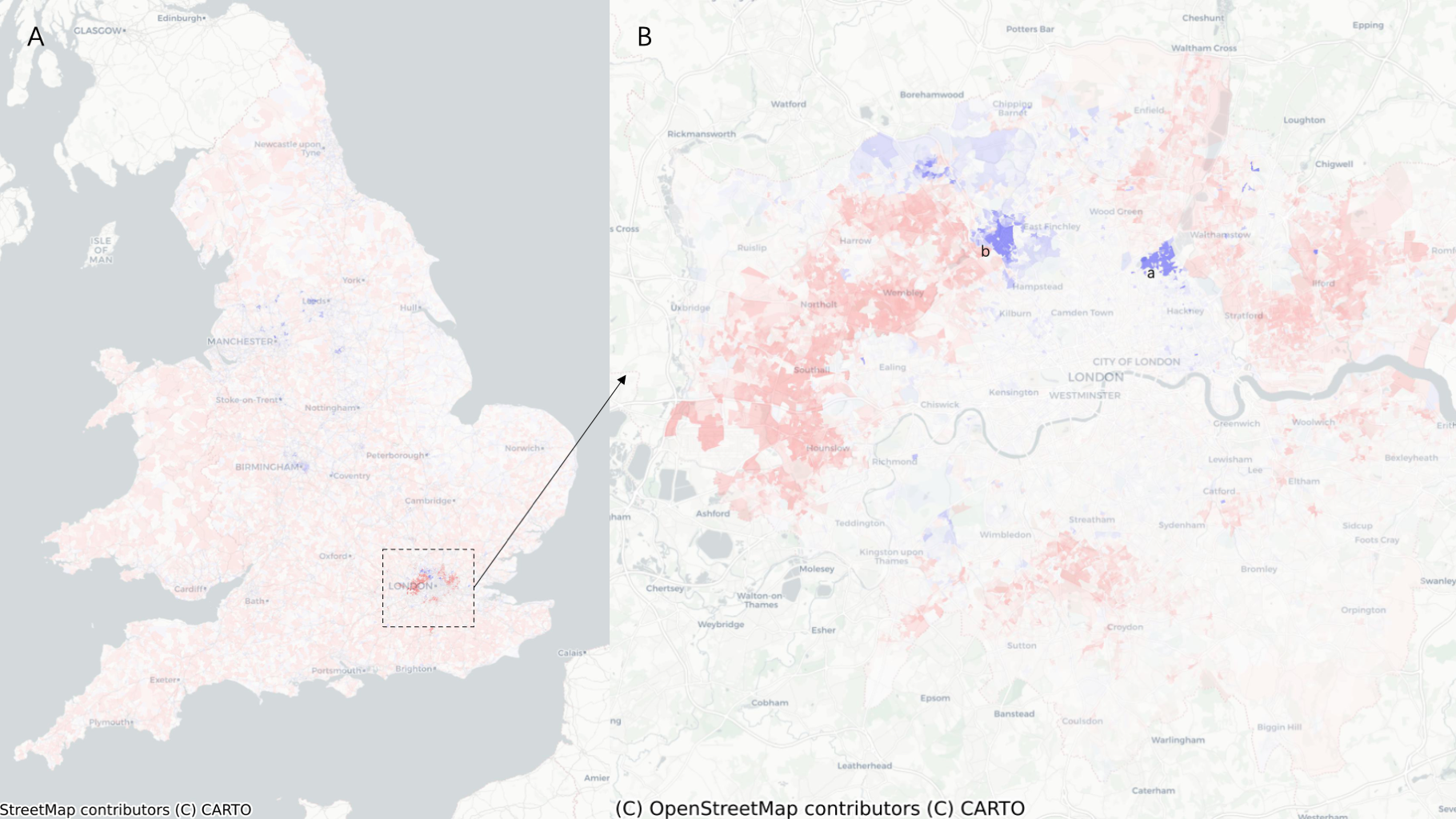}
    \caption{The spatial clustering of Tamil related people found by $\eta_{16}$ of England and Wales (\textbf{A}) and London (\textbf{B}). Here \textbf{a} is Tamil Community Housing Association, and \textbf{b} is Tamil Association of Brent. The darkest red region are however the clusters of Yiddish and Hebrew speakers.}
    \label{fig:tamil}
\end{figure}

The third highest localized feature in London, $\eta_{16}$, is associated with social variables related to the Tamil community and Yiddish, Israeli, and Hebrew speakers. Negative entries of $\eta_{16}$ indicate the presence of the Tamil community near the Tamil Community Housing Association, which supports refugees from Sri Lanka. The Tamil community in London has been growing since the Sri Lankan Civil War and is becoming distinct, as evidenced by high academic performance of Tamil children and a preference for having only children. Meanwhile, positive entries of $\eta_{16}$ mark areas with high concentrations of Yiddish, Israeli, and Hebrew speakers in Stamford Hill, North London. These areas tend to be isolated, as seen in the distribution of Yiddish newspapers aimed at audiences in Leeds, Manchester, and Gateshead, rather than being clustered in a distinct Yiddish neighborhood.

The two eigenvectors, $\eta_2$ and $\eta_{11}$, provide insight into the occupational landscape of London. $\eta_2$ is highly correlated with areas that demand degrees and higher education, such as BA, BSc, MA, PhD, and PGCE, with a correlation coefficient of 0.85. This indicates a strong presence of professional and highly educated individuals in these areas. On the other hand, $\eta_{11}$ marks communities with a higher concentration of lower supervisory and technical occupations, including mechanics, chefs, train drivers, plumbers, and electricians, with a correlation coefficient of 0.33. These are typically considered higher grade blue-collar jobs that require specialized skills.

It is worth noting that $\eta_2$ also has a negative correlation with South Asian language speakers, specifically those who speak Pakistani Pahari, Mirpuri, and Potwari, indicating a lack of assimilation into London's societies. This may suggest a potential barrier for these individuals in accessing higher education and professional opportunities.

\begin{figure}[ht]
    \centering
    \includegraphics[width=0.99\linewidth]{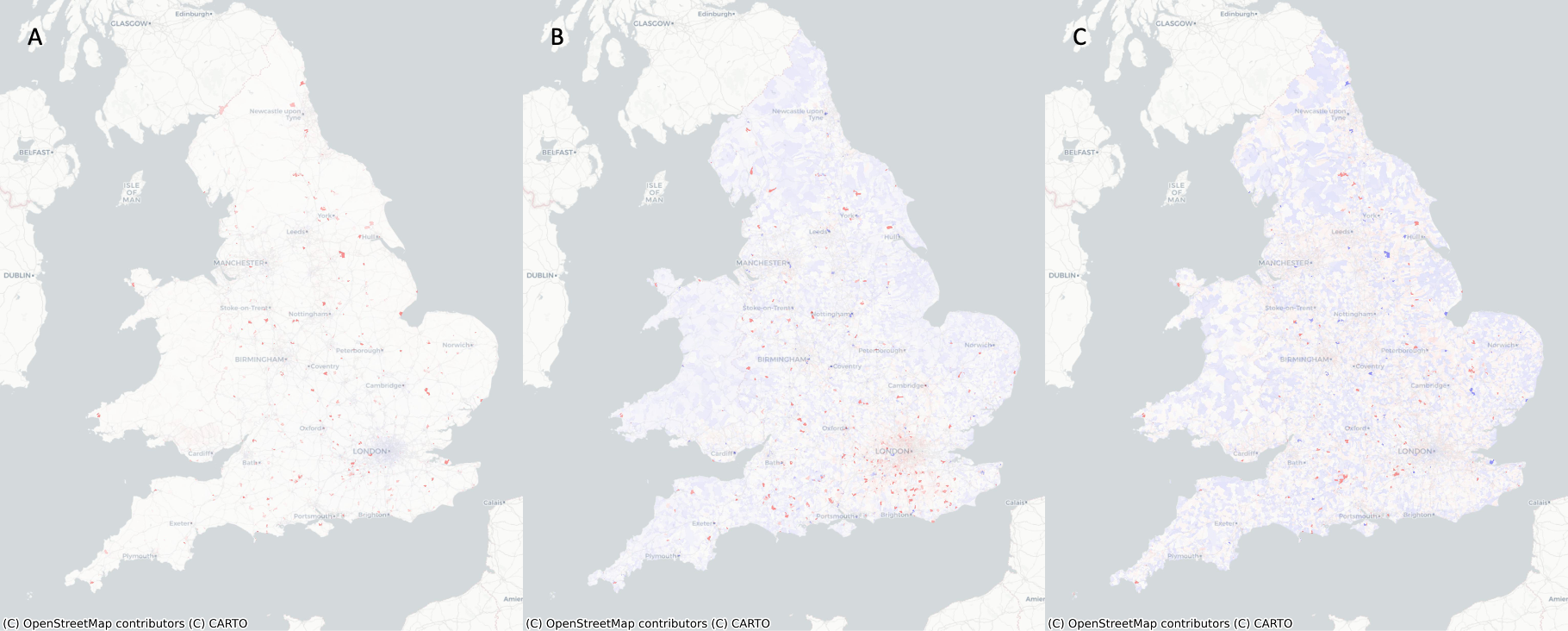}
    \caption{Three of the most globalized eigenvectors: \textbf{A}. $\eta_3$ maps prison installations; \textbf{B}. $\eta_8$ maps educations before college; and \textbf{C}. $\eta_4$ highlights the military camps.}
    \label{fig:global}
\end{figure}

The eigenvectors identified by $\eta_3$, which highlight prison installations~(Fig.~\ref{fig:global}A), are not unique to London, but can also be found in other cities. This eigenvector has the highest level of globalization among the first 20 (with a correlation coefficient of 0.010373), indicating that these features may be associated with broader infrastructure elements, such as $\eta_8$ for education~(Fig.~\ref{fig:global}B) or $\eta_4$ for national defense~(Fig.~\ref{fig:global}C). This highlights the importance of considering the broader contextual factors that influence local patterns and structures in cities, beyond just their specific local features.

\section{Discussion}\label{sec12}

In this article, we applied diffusion maps to analyze the synchronized variations in the census responses of England and Wales. Our study represents a novel attempt to decompose the British census as a whole, not just in urban areas. The results of our work demonstrate the effectiveness of diffusion maps in uncovering the underlying social structures in bulk, publicly accessible data. Our method ranks the relative importance of different features by themes and highlights the continuity of social aspects, such as educational levels, in the form of continuous indices.

The complex nature of demographic features calls for a efficient and scalable data analysis approach that can handle multiple scales and themes. Manifold learning methods, such as diffusion maps, are ideal for this purpose as they focus on local structures while preserving global information. To adapt the diffusion maps method to the bulk census data of England and Wales, we developed several techniques, including the preservation of a limited number of correlations that ensure connectedness and the use of Spearman Rank Correlation to measure the high-dimensional census data and account for heterogeneity in the distribution of social variables. 

Networks with a higher likelihood of power law behaviors in the degree distribution have robust cross-scale characteristics, and thus defining local metrics can better reflect the heterogeneous distribution of their data noise using local metrics. Diffusion mapping can portray nonlinear features and local metrics in the data, thus one single model can be used to deal with large cross-scale problems that reveals both general picture and local hotspots. 

The diffusion mapping eigenvectors provide insights into the urban structures of England and Wales and their impact on the cross-scaled behaviors of British society. Our method uses a heuristic definition of the $k$ nearest neighbor network to ensure that these characteristics are globally sensible and applicable to all areas, not just cities. Furthermore, the local inverse participation ratio is used to identify minority groups localized in big cities and to classify features as sublinear or superlinear urban indicators using only one input dataset.'

The proposed local inverse participation ratio ($LIPR$) is a method for identifying and characterizing minorities in urban areas. The $LIPR$ measures the concentration of a given feature in a specific region compared to its distribution across the entire urban area. It calculates the fraction of total variation in a feature that is captured by a limited number of Output Areas. The $LIPR$ values of each feature allow us to categorize the features as either sublinear or superlinear urban indicators. Features with high $LIPR$ values are considered highly concentrated in one region and classified as superlinear indicators, while features with low $LIPR$ values are considered widely distributed and classified as sublinear indicators.

Our findings extend the existing knowledge that some urban indicators, such as the number of university students, can be infrastructural in some cities but are urban outputs in the others. The $LIPR$ provides valuable insights into the distributional patterns of demographic features in urban areas and can reveal the unique social, economic, and cultural characteristics of highly concentrated minority groups and their relationship to the broader urban population.

%

\subsection*{Acknowledgement}

We thank Professor Thilo Gross for providing the idea and the original work that understand cities through manifold learning. The authors thank Yiyun Liang for the fruitful discussion about the features interpretations for London. 

\newpage

\section*{Supplementary Materials}

\subsection*{Correlation Tables}

\begin{table*}[ht]
    \begin{tabular}{|l|r|}
        \hline
    Living in a couple: Married or in a registered same-sex civil partnership  & 0.82070 \\
    \hline
    2 cars or vans in household                                                & 0.80857 \\
    \hline
    Married                                                                    & 0.78467 \\
    \hline
    Occupancy rating (rooms) of +2 or more                                     & 0.78416 \\
    \hline
    2 cars or vans in household                                                & 0.77907 \\
    \hline
    All categories: Car or van availability                                    & 0.77101 \\
    \hline
    All categories: Car or van availability                                    & 0.77101 \\
    \hline
    Married couple household: No dependent children                            & 0.76956 \\
    \hline
    One family only: Married or same-sex civil partnership couple: No children & 0.76688 \\
    \hline
    Owned: Owned outright                                                      & 0.74956 \\
    \hline
    Whole house or bungalow: Detached                                          & 0.74835 \\
    \hline
    Unshared dwelling: Whole house or bungalow: Detached                       & 0.74312 \\
    \hline
    Median age                                                                 & 0.73191 \\
    \hline
    Up to 0.5 persons per room                                                 & 0.72897 \\
    \hline
    3 cars or vans in household                                                & 0.72776 \\
    \hline
    One family only: Married couple: No children                               & 0.72671 \\
    \hline
    Owned: Total                                                               & 0.72349 \\
    \hline
    One family only: Married or same-sex civil partnership couple: No children & 0.72308 \\
    \hline
    Owned: Owned outright                                                      & 0.72245 \\
    \hline
    3 cars or vans in household                                                & 0.72103 \\
    \hline
    Unshared dwelling: Whole house or bungalow: Detached                       & 0.71489 \\
    \hline
    One family only: All aged 65 and over                                      & 0.70880 \\
    \hline
    Owned: Owned outright                                                      & 0.69474 \\
    \hline
    One family only: All aged 65 and over                                      & 0.69380 \\
    \hline
    Age 55 to 64: Two or more person household: No dependent children          & 0.69327 \\ \hline
    \end{tabular}
    \caption{The most correlated census social variables with $\eta_1$.}
\end{table*}

\begin{table*}[ht]
    \begin{tabular}{|l|r|}
        \hline
    Degree (for example BA, BSc), Higher degree (for example MA, PhD, PGCE) & 0.84546 \\ \hline 
    2+ A levels/VCEs, 4+ AS levels, Higher School Certificate, Progression/Advanced Diploma, &\\ Welsh Baccalaureate Advanced Diploma & 0.80772 \\ \hline
    Highest level of qualification: Level 4 qualifications and above & 0.80735 \\ \hline
    1.2 Higher professional occupations & 0.75042 \\ \hline
    1. Higher managerial, administrative and professional occupations & 0.73521 \\ \hline
    Professional occupations; Business, media and public service professionals & 0.72533 \\ \hline
    2. Professional occupations & 0.71416 \\ \hline
    M Professional, scientific and technical activities & 0.70869 \\ \hline
    L3.1 Traditional employees & 0.70769 \\ \hline
    5+ O level (Passes)/CSEs (Grade 1)/GCSEs (Grades A*-C), School Certificate, & \\ 1 A level/2-3 AS levels/VCEs, Higher Diploma, Welsh Baccalaureate Intermediate Diploma & 0.69515 \\ \hline
    Associate professional and technical occupations  & 0.68851 \\ \hline
    L4 Lower professional and higher technical occupations & 0.68286 \\ \hline
    J Information and communication & 0.66708 \\ \hline
    Day-to-day activities not limited & 0.66264 \\ \hline
    2. Lower managerial, administrative and professional occupations & 0.65226 \\ \hline
    L3.2 New employees & 0.63474 \\ \hline
    Professional occupations; Science, research, engineering and technology professionals & 0.63435 \\ \hline
    Very good health & 0.63065 \\ \hline
    Professional qualifications (for example teaching, nursing, accountancy) & 0.61972 \\ \hline
    3. Associate professional and technical occupations & 0.61193 \\ \hline
    Different ethnic groups within partnerships & \\(whether or not different ethnic groups between generations) & 0.60756 \\ \hline
    L4.1 Traditional employees & 0.60653 \\ \hline
    Foreign qualifications & 0.60465 \\ \hline
    Professional occupations; Science, research, engineering and technology professionals; & \\ Information Technology and Telecommunications Professionals & 0.60303 \\ \hline
    British only identity & 0.60164
    \\ \hline
    \end{tabular}
    \caption{The most correlated census social variables with $\eta_2$.}
    \end{table*}

\end{document}